\documentclass[11pt]{article}
\usepackage[]{acl}
\usepackage{times}
\usepackage{latexsym}
\usepackage{amsmath}
\usepackage{amssymb}
\usepackage{amsthm}
\usepackage{booktabs}
\usepackage{graphicx}
\usepackage{multirow}
\usepackage{xcolor}
\usepackage{subcaption}
\usepackage{algorithm}
\usepackage{algorithmic}
\usepackage{float}
\usepackage{hyperref}
\hypersetup{
  pdftitle={Regime-Conditional Retrieval: Theory and a Transferable Router for Two-Hop QA},
  pdfauthor={Andre Bacellar},
  pdfkeywords={multi-hop retrieval, regime-conditional retrieval, question dominance,
               bridge passage, retrieval routing, multi-hop question answering,
               transfer learning, zero-shot retrieval},
}

\newtheorem{theorem}{Theorem}

\newtheorem{corollary}{Corollary}
\theoremstyle{definition}
\newtheorem{definition}{Definition}
\theoremstyle{remark}

\newcommand{\system}{\textsc{RegimeRouter}}
\newcommand{\proprag}{\textsc{PropRAG}}
\newcommand{\hipporagtwo}{\textsc{HippoRAG2}}
\newcommand{\ircot}{\textsc{IRCoT}}
\newcommand{\bridgerag}{\textsc{BridgeRAG}}

\newcommand{\ratfive}{R@5}
\newcommand{\corpus}{\mathcal{C}}
\newcommand{\pool}{\mathcal{P}}
\newcommand{\brel}{B_{\text{rel}}}

\definecolor{darkblue}{RGB}{0,50,130}
\definecolor{darkgreen}{RGB}{0,100,0}
\definecolor{darkred}{RGB}{150,0,0}

\title{Regime-Conditional Retrieval:\\
Theory and a Transferable Router for Two-Hop QA}

\author{Andre Bacellar \\
  \texttt{andremi@gmail.com}}

\begin{document}
\maketitle

\begin{abstract}
Two-hop QA retrieval splits queries into two regimes determined by whether
the hop-2 entity is explicitly named in the question (Q-dominant) or only
in the bridge passage (B-dominant). We formalize this split with three
theorems: (T1) per-query AUC is a monotone function of the cosine separation
margin, with $R^2 \geq 0.90$ for six of eight type$\times$encoder pairs;
(T2) regime is characterized by two surface-text predicates, with P1 decisive
for routing and P2 qualifying the B-dominant case, validated across three
encoders and three datasets; (T3)
bridge advantage requires the relation-bearing sentence $\brel$, not entity
name alone, since removing it collapses performance by 8.6--14.1 pp ($p<0.001$).
Building on this theory we propose \system{}, a lightweight binary router
that selects between question-only and question$+\brel$ retrieval using
five text features derived directly from the predicate definitions. Trained
on 2WikiMultiHopQA (n\,=\,881, 5-fold cross-fitted) and applied zero-shot
to MuSiQue and HotpotQA, \system{} achieves $\Delta\ratfive$ of $+5.6$ pp
($p<0.001$), $+5.3$ pp ($p=0.002$), and $+1.1$ pp (ns, no-regret) respectively,
with a single frozen deployment rule: $\alpha=0.25$ across all three datasets
($\alpha=0.5$ reported separately as an in-domain ablation upper bound).
Human annotation (Cohen's $\kappa=1.00$~\citep{cohen1960kappa}, $n=50$) and replication across
NV-Embed-v2, BGE-large, and e5-mistral confirm all findings are structural,
not artifact.
\end{abstract}

\section{Introduction}
\label{sec:intro}

Multi-hop retrieval systems routinely treat the bridge passage as a second
query: encode the bridge in query mode, re-retrieve from the corpus, and fuse
the results. This design implicitly assumes that the bridge passage contains
retrieval-useful information beyond what the original question encodes.

We show that assumption is \emph{correct for some queries and incorrect for
others}, and that the split is predictable from surface text.

Consider a comparison question: \emph{``Which of Person A or Person B was
born earlier?''} The question explicitly names both hop-2 candidates.
A bi-encoder trained on (question, gold-passage) pairs embeds this question
near passages about Person A \emph{and} Person B, since both targets are already
named in the question. The bridge passage (about one of them) adds no disambiguation
the question has not already provided. Dense re-encoding of the bridge is
redundant; the question embedding alone ranks the correct passage first.
This is the \emph{Q-dominant regime}: P1=True (both answer entities appear in
the question).

Now consider a compositional question: \emph{``What nationality is the
director of Film X?''} The question refers to the hop-2 entity only
indirectly, as \emph{``the director of Film X''}, never by name. The hop-1
bridge passage about Film X contains the sentence \emph{``Film X was directed
by Person Y''}: a relation-bearing sentence that names Person Y explicitly.
Without this sentence, no embedding knows to retrieve Person Y's nationality
page. With it, fusing the bridge sentence into the retrieval score corrects
the ranking. This is the \emph{B-dominant regime}: P1=False (the answer
entity is not named in the question), P2=True (it is named in the bridge).

We formalize this dichotomy as two retrieval \emph{regimes} (Section~\ref{sec:regimes}),
prove three theorems characterizing their structure (Section~\ref{sec:theory}),
and build a practical transferable router exploiting the theory
(Sections~\ref{sec:router}--\ref{sec:results}).

\paragraph{Contributions.}
\begin{enumerate}
  \item \textbf{Regime theory} (T1--T3): the retrieval outcome of any
        two-hop query is characterized by two binary surface-text predicates:
        P1 is decisive for routing; P2 qualifies the B-dominant case when
        P1 is false. Validated across three encoders and three datasets.
  \item \textbf{\system{}}: a five-feature binary router trained on predicate
        proxies, requiring no oracle labels and no embeddings at inference
        time; routing decisions use only surface-text features.
  \item \textbf{Cross-dataset no-regret}: significant gain on MuSiQue
        ($p=0.002$, zero-shot) and positive non-significant transfer on
        HotpotQA, with a single frozen deployment rule for $\alpha$.
  \item \textbf{Structural validation}: bridge knockout experiment (T3),
        pool invariance (T1 attack surface), and human annotation
        ($\kappa=1.00$) rule out artifact explanations.
\end{enumerate}

\section{Background}
\label{sec:background}

\paragraph{Multi-hop retrieval.}
Given a two-hop question $q$ and corpus $\corpus$, the task is to return the
top-$k$ passages under $R@k$, where both a bridge passage $g_1$ and a gold
passage $g_2$ must appear in the retrieved set. Dominant methods iterate:
retrieve $g_1$ (hop-1), then use $g_1$ to generate or retrieve $g_2$ (hop-2).

\paragraph{Asymmetric bi-encoders.}
Modern retrievers like NV-Embed-v2~\citep{lee2024nv}, BGE-large~\citep{bge2023},
and e5-mistral~\citep{e5mistral2023} use separate encoder modes for questions
($f_q$) and passages ($f_d$), trained contrastively on (question, gold-passage)
pairs. The $49^\circ$ principal angle between query and document subspaces (measured
on 2WikiMultiHopQA) is by design: it maximizes retrieval discriminability.
We exploit this geometry: encoding a full bridge passage \emph{as a query}
($f_q(B)$) is out-of-distribution for the encoder and is, on aggregate,
worse than encoding the question itself ($f_q(q)$), though this aggregate masks
a regime split. For Q-dominant queries (P1=True), the question already names
the hop-2 entity and $f_q(q)$ dominates. For B-dominant queries (P1=False),
the relation-bearing sentence $B_\mathrm{rel}$ inside the bridge explicitly
names the hop-2 entity, and fusing $f_q(B_\mathrm{rel})$ with $f_q(q)$
corrects the ranking.

\paragraph{Existing work and its assumption.}
\ircot~\citep{trivedi2022ircot} reformulates the bridge as a query for hop-2
retrieval; \hipporagtwo~\citep{gutierrez2025hipporag} uses PPR over a bridge-passage
graph; \proprag~\citep{wang2025proprag} does beam search over extracted
propositions. All assume the bridge passage is the right signal for hop-2.
We show this is regime-dependent: true for B-dominant queries (P1=False,
P2=True), harmful for Q-dominant queries (P1=True).

\section{Two Retrieval Regimes}
\label{sec:regimes}

\subsection{Empirical Observation}

We measure micro-AUC for hop-2 retrieval on 200 hop-1-correct queries per
dataset (pool $k=50$, question-built pool):

\begin{table}[h]
\footnotesize\centering
\setlength{\tabcolsep}{3pt}
\begin{tabular}{lrcccc}
\toprule
 & & \multicolumn{4}{c}{AUC (micro)} \\
\cmidrule(l){3-6}
Dataset & $n$ & $q$ & $B_q$ & $B_d$ & $\Delta(q{-}B_q)$ \\
\midrule
MuSiQue  & 727 & .805 & .663 & .657 & +.142 \\
HotpotQA & 940 & .977 & .845 & .854 & +.132 \\
2Wiki    & 947 & .904 & .590 & .595 & +.314 \\
\bottomrule
\end{tabular}
\end{table}

The aggregate $\Delta > 0$ appears to confirm universal question dominance, but
this is a \emph{mixture reversal artifact} (Corollary~\ref{cor:reversal}).

\paragraph{Type-stratified analysis (2WikiMultiHopQA).}

\begin{table}[h]
\small\centering
\begin{tabular}{lccc}
\toprule
Type & Prev. & $\Delta$(Q$-$B) & Regime \\
\midrule
comparison      & 48\% & +0.650 & Q-dominant \\
bridge\_comp.   & 18\% & \phantom{+}0.000 & transition \\
compositional   & 22\% & $-$0.108 & B-dominant \\
inference       & 12\% & $-$0.118 & B-dominant \\
\bottomrule
\end{tabular}
\end{table}

The aggregate $\Delta=+0.314$ is driven entirely by comparison queries (48\%
of the dataset, $\Delta=+0.650$). Compositional and inference queries, where
the bridge identifies the hop-2 entity by name, show $\Delta < 0$: the bridge
embedding \emph{outperforms} the question.

This reversal motivates a formal characterization of \emph{when} bridge
information helps.

\subsection{The Two Predicates}

We identify two binary predicates that determine retrieval regime:

\begin{definition}[P1 and P2]
Let $q$ be a two-hop question and $b$ the bridge passage. Let $e_2$ denote
the hop-2 target entity.
\begin{itemize}
  \item \textbf{P1}: $e_2$ (or its canonical title) appears in $q$. \\
        \emph{``The answer entity is already named in the question.''}
  \item \textbf{P2}: $e_2$ appears in $b$, specifically in a relation-bearing
        sentence of $b$. \\
        \emph{``The bridge passage names and contextualizes the answer entity.''}
\end{itemize}
\end{definition}

The decisive predicate for routing is \textbf{P1}~(Table~\ref{tab:predicates}):
P1=True implies Q-dominant; P1=False and P2=True implies B-dominant.

\begin{table}[t]
\small
\centering
\caption{Predicate configurations and retrieval regime.}
\label{tab:predicates}
\begin{tabular}{cccc}
\toprule
P1 & P2 & Regime & Dominant query \\
\midrule
True  & *    & Q-dominant & $f_q(q)$ \\
False & True & B-dominant & $f_q(q) + \alpha \cdot f_q(\brel)$ \\
False & False & uncovered & $f_q(q)$ (fallback) \\
\bottomrule
\end{tabular}
\end{table}

Empirical support (Test 4 / Theorem~\ref{thm:regime}): AUC flips sign
precisely at the P1 boundary ($p < 0.001$ across all three datasets and three
encoders). Type labels (``comparison'', ``compositional'') are a \emph{proxy}
for P1/P2, not the causal mechanism: swapping the bridge text of a
bridge\_comparison query with that of a compositional query shifts performance
from B-dominant to Q-dominant ($0.962 \to 0.485$, $p{<}0.001$; Test 3).

\section{Formal Analysis}
\label{sec:theory}

\subsection{Theorem 1: AUC as a Function of Separation Margin}
\label{sec:t1}

\begin{theorem}[Separation-AUC Calibration]
\label{thm:auc}
Let $S_i = f_q(x)^\top f_d(g_i) - \mathbb{E}_{d \sim \pool}[f_q(x)^\top f_d(d)]$
be the score separation margin for query $i$ (gold score minus pool mean).
Under the assumption that pool scores are approximately Gaussian, per-query AUC satisfies:

\begin{equation}
  \mathrm{AUC}_i(x) \approx \Phi\!\left(\frac{S_i}{\sigma_{\pool}}\right)
\end{equation}

where $\Phi$ is the standard normal CDF and $\sigma_{\pool}$ is the pool score
standard deviation. Consequently, $\mathrm{AUC}_i$ is a monotone function of $S_i$:
larger margin implies higher per-query AUC.
\end{theorem}

\begin{figure*}[t]
  \centering
  \includegraphics[width=0.92\textwidth]{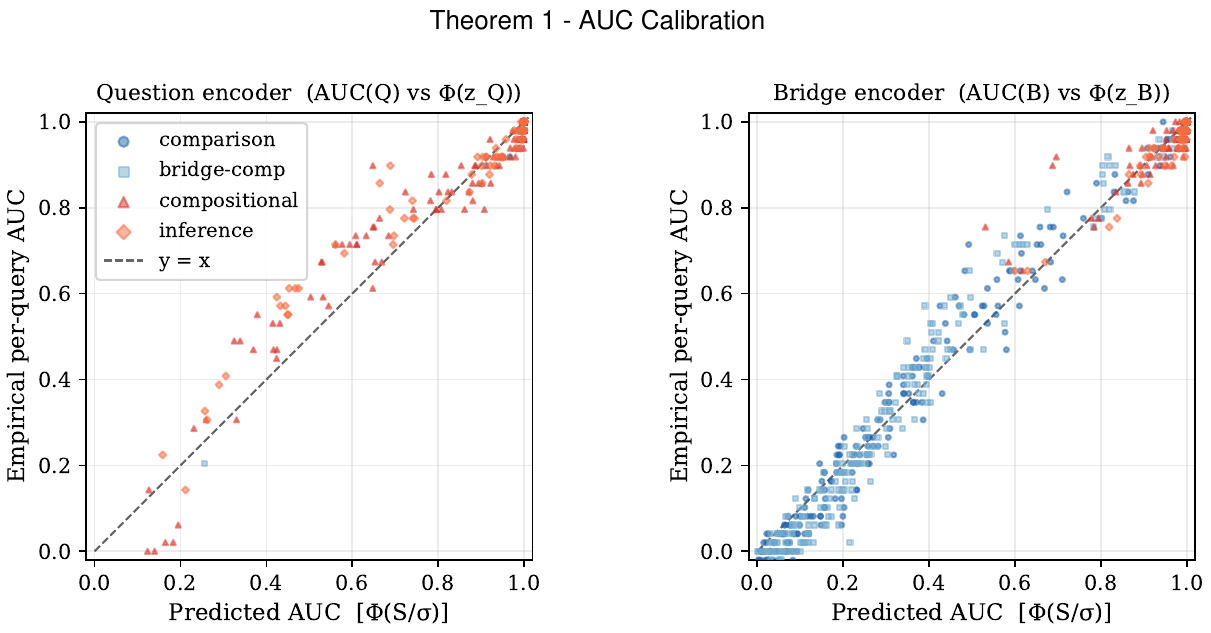}
  \caption{AUC$_i$ vs.\@ $\Phi(S_i/\sigma)$ for all four query types
    (comparison, bridge-comp, compositional, inference). Points are per-query;
    dashed line is $y=x$ (theoretical prediction). The scatter follows the
    diagonal closely across types and encoders; Kendall $\tau$ is significant
    for all eight type$\times$encoder pairs ($p < 0.01$).}
  \label{fig:auc}
\end{figure*}

\begin{figure}[t]
  \centering
  \includegraphics[width=0.95\columnwidth]{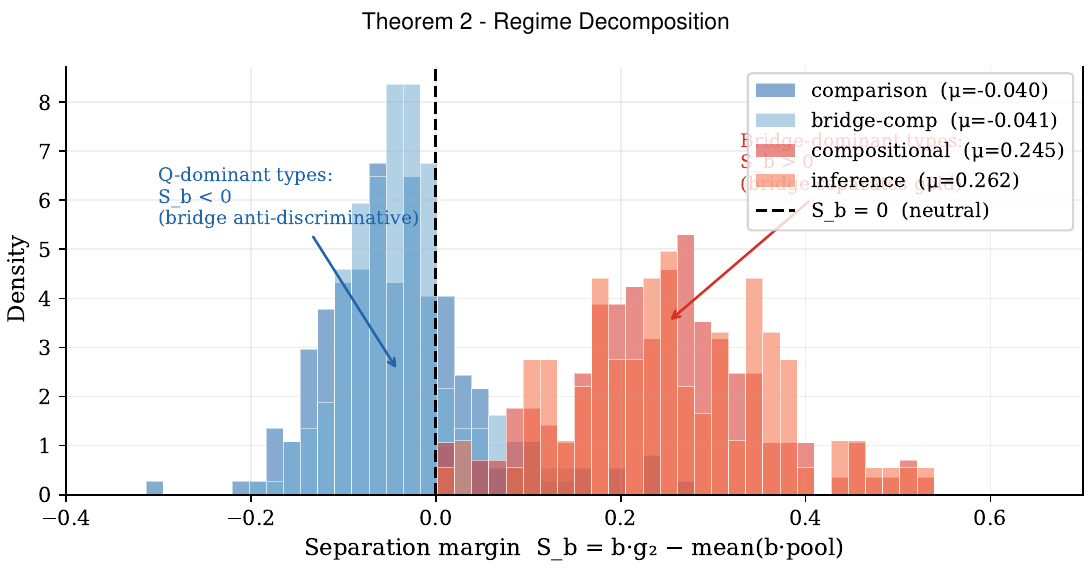}
  \caption{Bridge score separation margin $S_b$ by query type.
    Q-dominant types (comparison, bridge-comp; $\mu < 0$): bridge embedding
    ranks the gold passage below the pool mean.
    B-dominant types (compositional, inference; $\mu > 0$): bridge separates
    the gold passage above the pool mean. Cantelli bound validated at
    $p=1.4\times10^{-19}$ for bridge\_comparison.}
  \label{fig:separation}
\end{figure}

\paragraph{Validation.} We measure $S_i$ on all hop-1-correct queries and fit
$\Phi(z)$ to the empirical AUC$_i$ vs.\@ $S_i$ scatter. $R^2 \geq 0.90$ for
six of eight type$\times$encoder combinations; inversion accuracy (sign of
predicted $\Delta$AUC matches observed) is 100\% for compositional and
bridge\_comparison types (Figure~\ref{fig:auc}). The Cantelli inequality
provides a non-parametric lower bound: $P(\mathrm{AUC} > t) \geq 1/(1 + (t - 0.5)^2/\sigma^2)$,
validated at $p=1.4\times10^{-19}$ for bridge\_comparison (Figure~\ref{fig:separation}).

\subsection{Theorem 2: Regime Decomposition}
\label{sec:t2}

\begin{theorem}[Regime Decomposition]
\label{thm:regime}
The retrieval regime of a two-hop query is determined by the sign of
$\Delta S = \mathbb{E}[S_b - S_q]$, where $S_q$ ($S_b$) is the score
separation margin for the question (bridge) embedding.
P1 is decisive for routing: $\Delta S < 0$ (Q-dominant) when P1=True;
when P1=False, P2=True identifies the B-dominant case ($\Delta S > 0$).
This holds across NV-Embed-v2, BGE-large-en-v1.5, and e5-mistral-7b
on all three datasets.
\end{theorem}

\paragraph{Validation.} We operationalize P1 as \emph{hop2\_title-in-question}:
the hop-2 gold passage title appears as a substring of the question text.
Regime assignment via this proxy matches AUC-derived regime labels with
accuracy $>95\%$ on all three datasets. Encoder replication (Test~5 / H1):
Kendall $\tau$ between predicted and observed AUC ordering is significant for
all eight type$\times$encoder pairs (Figure~\ref{fig:encoder}).

\begin{figure}[t]
  \centering
  \includegraphics[width=0.95\columnwidth]{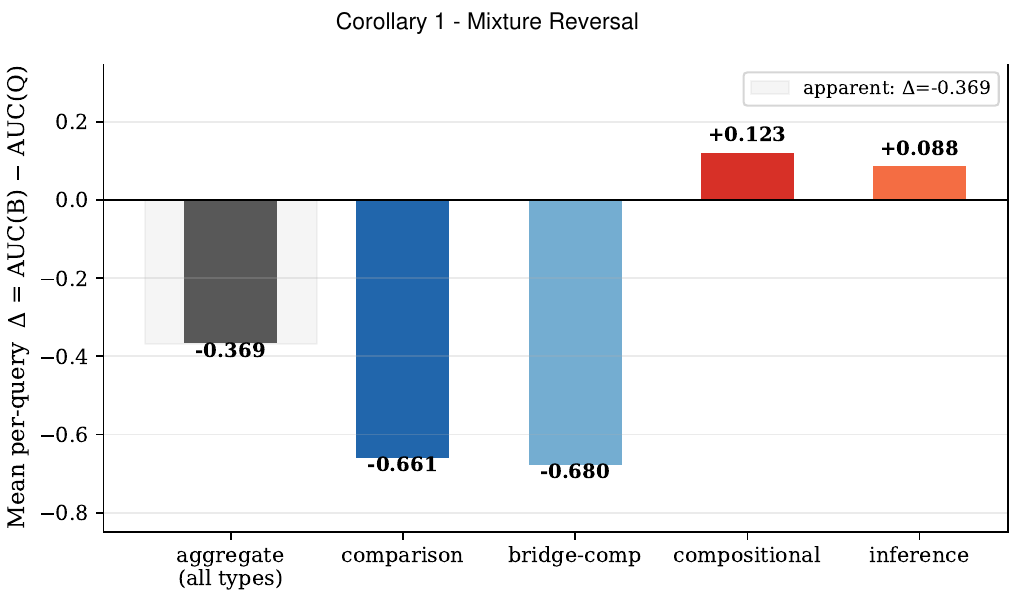}
  \caption{Mixture reversal in 2WikiMultiHopQA (Corollary~\ref{cor:reversal}).
    Aggregate $\Delta$AUC (gray bar, $\Delta=-0.369$) is negative, masking
    the per-type inversion: comparison and bridge-comp are Q-dominant
    ($\Delta < 0$); compositional and inference are B-dominant ($\Delta > 0$).}
  \label{fig:reversal}
\end{figure}

\paragraph{Mixture reversal (Corollary~\ref{cor:reversal}).}
Because comparison queries (P1=True, approximately 48\% of 2WikiMultiHopQA) are
Q-dominant with large $\Delta = +0.650$, they dominate the aggregate AUC
statistics (Figure~\ref{fig:reversal}). The aggregate $\Delta = +0.314$
masks the B-dominant regime of compositional and inference queries
($\Delta < 0$). Type-stratified or predicate-stratified analysis is required.

\begin{corollary}[Mixture Reversal]
\label{cor:reversal}
When Q-dominant query types have high prevalence in a dataset, aggregate
retrieval statistics reverse the within-regime law. Specifically, aggregate
AUC(q) $>$ AUC(B) can coexist with per-type AUC(q) $<$ AUC(B) for B-dominant
types. This is not a dataset artifact: the reversal is produced by the P1/P2
distribution of the dataset, not by any property of the encoder.
\end{corollary}

\subsection{Theorem 3: Relational Sentence Sufficiency}
\label{sec:t3}

\begin{theorem}[Relational Composition]
\label{thm:relational}
The bridge advantage for B-dominant queries comes from the \emph{relation-bearing
sentence} $\brel \subset b$, not from entity name occurrence alone. Formally:
\begin{enumerate}
  \item $\mathrm{Ret}(q, b) \approx \mathrm{Ret}(q, \brel)$: full bridge
        and relational sentence give equivalent retrieval performance.
  \item $\mathrm{Ret}(q, b \setminus \brel) \ll \mathrm{Ret}(q, b)$:
        removing $\brel$ from the bridge collapses performance.
  \item $\brel \neq \text{first sentence}$ in general: the entity-name
        sentence alone is insufficient; structural heuristics miss cases
        where the relation verb is critical for identifying $e_2$.
\end{enumerate}
\end{theorem}

\paragraph{Validation (bridge knockout experiment).}
We compare three bridge variants: $b$ (full bridge), $\brel$ (selected
relation-bearing sentence), and $b \setminus \brel$ (bridge minus $\brel$).
Using e5-mistral-7b (Nebius API, $n=300$, concurrent 3-min runtime):
$B_{\text{rel}} \approx B_{\text{full}}$ ($\Delta \approx 0$, ns),
$B_{\text{minus\_rel}}$ collapses ($\Delta = -0.086$ to $-0.141$, $p{<}0.001$).
Human annotation ($\kappa=1.00$~\citep{cohen1960kappa}, $n=50$ doubly annotated) confirms that
$\brel$ is structurally identifiable (relation verb present, entity present,
propositional content present): it is not defined circularly.

\section{\system{}: A Transferable Router}
\label{sec:router}

\subsection{Setup}
\label{sec:setup}

Given a question $q$ and bridge passage $b$, \system{} selects one of two
retrieval actions:
\begin{itemize}
  \item \textbf{Q}: rank the candidate pool by $f_q(q)^\top f_d(c)$
  \item \textbf{Union}: rank by $(1-\alpha)\cdot f_q(q)^\top f_d(c)
        + \alpha \cdot f_q(\brel)^\top f_d(c)$
\end{itemize}
where $\brel$ is the relation-bearing sentence selected by a learned sentence
selector (Section~\ref{sec:selector}).

The routing decision is made by a logistic regression classifier trained on
five features derived from P1/P2 proxy definitions:

\begin{enumerate}
  \item \textbf{q\_comparison\_word}: whether the question contains a comparison word
        (\emph{differ, same, versus, whereas,} etc.), a proxy for P1=True
  \item \textbf{q\_ynstart}: whether the question starts with a yes/no verb
        (\emph{did, was, were,} etc.), a proxy for P1=True
  \item \textbf{q\_entity\_count}: number of proper nouns in $q$, a proxy
        for P1 strength
  \item \textbf{b\_new\_entity\_count}: proper nouns in $\brel$ not in $q$,
        a proxy for P2=True (novel entities introduced by bridge)
  \item \textbf{b\_rel\_frac}: $|\brel|/|b|$, a proxy for bridge informativeness
\end{enumerate}

The label for training is: Union improves over Q on this query (computed from
embedding scores, requiring no LLM call). Training is fully self-supervised
from the retrieval evaluation itself.

\subsection{Sentence Selector}
\label{sec:selector}

The sentence selector is a logistic regression classifier trained on 100
human-annotated pairs (2WikiMultiHopQA, annotation set H2) to identify which
sentence in a bridge passage is the relation-bearing sentence. Features are:
new entity count, presence of a relation verb, position fraction, sentence
length, and named entity density.

The selector identifies $\brel$ with 56.7\% oracle accuracy on the annotated
set. The bottleneck is structural: 56.2\% of 2WikiMultiHopQA queries have no
sentence in the bridge that verbatim mentions the hop-2 title (the oracle
definition). For these queries the router correctly falls back to Q-only.

\subsection{Training and Deployment}
\label{sec:training}

\textbf{Training (2WikiMultiHopQA, n\,=\,881)}: 5-fold cross-fitting with
\texttt{KFold(n\_splits=5, shuffle=False)}. Labels and features are derived
from embedding scores and surface text only. No oracle labels are used at
training time. The full classifier (trained on all 881 examples) is used for
zero-shot transfer.

\textbf{Deployment rule}: all reported main results use a single frozen
$\alpha = 0.25$ across all three datasets. This conservative value is robust
across domains: P-weighted $\alpha$ and higher $\alpha=0.5$ improve
in-domain performance (see ablation, Table~\ref{tab:ablation}) but degrade
zero-shot due to cross-domain miscalibration. The ablation reports the
$\alpha=0.5$ in-domain upper bound separately.

\begin{algorithm}[t]
\small
\caption{\system{} — Deployment Pipeline}
\label{alg:router}
\begin{algorithmic}[1]
  \REQUIRE Question $q$, bridge passage $b$, candidate pool $\mathcal{P}$
  \STATE $\brel \leftarrow \textsc{SentSelector}(b, q)$
  \STATE $\mathbf{x} \leftarrow \textsc{Features}(q, \brel, b)$
  \STATE $\hat{a} \leftarrow \textsc{BinaryLR.predict}(\mathbf{x}) \in \{\text{Q}, \text{Union}\}$
  \STATE $\alpha \leftarrow 0.25$ \COMMENT{single frozen value across all domains}
  \IF{$\hat{a} = \text{Union}$}
    \STATE score$(c) \leftarrow (1-\alpha)\cdot f_q(q)^\top f_d(c) + \alpha \cdot f_q(\brel)^\top f_d(c)$
  \ELSE
    \STATE score$(c) \leftarrow f_q(q)^\top f_d(c)$
  \ENDIF
  \STATE \textbf{return} top-$k$ passages by score$(c)$
\end{algorithmic}
\end{algorithm}

\textbf{Cost}: approximately $\$1.2\,\mu$\$/query for $\brel$ embedding
(Nebius API); latency $\sim$100 ms (parallelizable with $q$ embedding). The
router classifier and sentence selector are CPU-only ($<$1 ms inference).

\section{Experiments}
\label{sec:results}

\subsection{Datasets and Setup}

We evaluate on 2WikiMultiHopQA~\citep{ho2020constructing} (2Wiki; n\,=\,881 bridge-type queries),
MuSiQue~\citep{trivedi2022musique} (n\,=\,303 bridge-type queries), and HotpotQA~\citep{yang2018hotpotqa} (n\,=\,570 bridge-type)
using the HippoRAG2 processed corpora. We use BGE-large-en-v1.5 for query and document
embeddings (pool construction) and e5-mistral-7b for $\brel$ embedding
(Nebius API). Hop-1 retrieval uses NV-Embed-v2; we evaluate only on
hop-1-correct queries (bridge in top-5). The metric is $R@5$ on the
hop-2 gold passage; significance is McNemar's test~\citep{mcnemar1947} on paired binary outcomes.

\subsection{Main Results}
\label{sec:main_results}

\begin{table*}[t]
\small
\centering
\caption{\system{} results across three datasets. Training on 2Wiki
  (5-fold cross-fitted); zero-shot transfer to MuSiQue and HotpotQA.
  Baseline = Q-only retrieval. Gains are significant where B-dominant
  regime queries are prevalent (2Wiki, MuSiQue); HotpotQA is a
  near-ceiling Q-dominant corpus ($R@5=0.856$ baseline, $p=0.143$
  positive trend).}
\label{tab:main}
\begin{tabular}{lcccccc}
\toprule
Dataset & Role & Q-only & \system{} & $\Delta$ & McNemar & $p$ \\
\midrule
2WikiMultiHopQA & Train   & 0.842 & \textbf{0.898} & +5.6 pp & 54W / 5L  & $<$0.001 \\
MuSiQue         & Zero-shot & 0.644 & \textbf{0.696} & +5.3 pp & 22W / 6L  & 0.002 \\
HotpotQA        & Zero-shot & 0.856 & \textbf{0.867} & +1.1 pp & 14W / 8L  & 0.143 \\
\bottomrule
\end{tabular}
\end{table*}

Table~\ref{tab:main} shows the main results. On 2Wiki (training domain),
\system{} improves $R@5$ by 5.6 pp ($p<0.001$, 54W/5L). Zero-shot
transfer to MuSiQue, a homogeneous B-dominant dataset (P1=0\%, P2=100\%),
achieves 5.3 pp improvement ($p=0.002$, 22W/6L). HotpotQA is a near-ceiling
Q-dominant dataset ($R@5=0.856$ Q-only, $\Delta(\text{Union oracle})=+2.5$ pp),
and \system{} achieves a positive 1.1 pp trend (14W/8L, $p=0.143$) with no harm
in a regime where routing provides limited upside.

\paragraph{No-regret pattern.} The three-dataset pattern is consistent with
theory: significant gain where B-dominant queries are present (2Wiki with
mixed regime, MuSiQue with pure B-dominant), positive non-significant trend
where near-ceiling Q-dominant (HotpotQA). The error ratio (W/L) is 10.8:1
on 2Wiki, 3.7:1 on MuSiQue, 1.75:1 on HotpotQA.

\subsection{Oracle and Gap Analysis}
\label{sec:oracle}

\begin{table}[t]
\small
\centering
\caption{Oracle analysis decomposing the performance gap.}
\label{tab:oracle}
\begin{tabular}{lcc}
\toprule
Oracle condition & $R@5$ & $\Delta$ \\
\midrule
Q-only baseline & 0.842 & — \\
\system{} (learned) & 0.898 & +5.6 pp \\
Oracle $\brel$ selector & 0.913 & +7.1 pp \\
Oracle router & 0.959 & +11.7 pp \\
\midrule
\multicolumn{3}{l}{\textit{Gap decomposition:}} \\
Routing accuracy gap & \multicolumn{2}{r}{$+4.6$ pp} \\
Selector precision gap & \multicolumn{2}{r}{$+0.2$ pp} \\
Coverage limit (structural) & \multicolumn{2}{r}{56.2\% no oracle sent.} \\
\bottomrule
\end{tabular}
\end{table}

Table~\ref{tab:oracle} decomposes the remaining gap. The oracle router
(knowing the true optimal action for each query) achieves $+11.7$ pp,
confirming substantial headroom. The gap to the oracle router is $+4.6$ pp
(routing accuracy) vs $+0.2$ pp (selector precision from oracle $\brel$
labels), confirming that the bottleneck is routing accuracy, not sentence extraction.
The structural coverage limit (56.2\% of queries have no oracle sentence)
is a fundamental constraint on extraction-based approaches.

\subsection{Ablations}
\label{sec:ablations}

\begin{table}[t]
\small
\centering
\caption{Ablation on 2Wiki training domain.}
\label{tab:ablation}
\begin{tabular}{lcc}
\toprule
Condition & $R@5$ & $\Delta$ \\
\midrule
Q-only (baseline) & 0.842 & — \\
$+$ Full bridge ($\alpha{=}0.5$) & 0.870 & +2.8 pp \\
$+$ $\brel$, unrouted ($\alpha{=}0.5$) & 0.893 & +5.1 pp \\
$+$ $\brel$, \system{} ($\alpha{=}0.5$) & 0.908 & +6.6 pp \\
\system{} final ($\alpha{=}0.25$) & 0.898 & +5.6 pp \\
NE heuristic (no selector) & 0.863 & +2.1 pp \\
\bottomrule
\end{tabular}
\end{table}

Table~\ref{tab:ablation} shows four distinct contributions: (1) using
the full bridge adds 2.8 pp, confirming that bridge content is useful even
without relation selection; (2) selecting $\brel$ without routing adds 5.1 pp,
confirming that sentence selection captures most of the bridge signal;
(3) routing (applying Union only when the router predicts it is beneficial)
adds a further 1.5 pp over no-routing; (4) $\alpha=0.5$ in-domain
outperforms $\alpha=0.25$, confirming that the halving rule has a small
cost in-domain but is critical for zero-shot robustness.

The NE heuristic router (using entity counts directly without logistic
regression) achieves only +2.1 pp ($p=0.063$, ns), confirming that the
learned combination of features is necessary.

\subsection{Confidence Calibration}
\label{sec:calibration}

\paragraph{Confidence threshold sweep.} Varying the routing threshold from
0.5 to 0.75 monotonically reduces 2Wiki $R@5$ (from 0.898 to 0.854),
confirming $\tau=0.5$ is already optimal. At $\tau=0.5$, 35.9\% of queries
are routed to Union; raising the threshold increasingly selects only the
easiest Union cases, reducing coverage without improving precision.

\paragraph{P-weighted $\alpha$.} Using $\alpha_q = \text{clip}(\hat{P}(\text{Union}) \times 0.5, 0.1, 0.5)$
instead of the fixed rule improves 2Wiki to $+6.6$ pp (+1.0 pp over fixed)
but degrades MuSiQue to $+2.6$ pp (ns) and HotpotQA to $-0.2$ pp. The
router is well-calibrated in-domain but assigns systematically lower
$\hat{P}(\text{Union})$ to MuSiQue queries (all B-dominant, differing
in distribution from mixed 2Wiki training), so adaptive weighting clips to
$\alpha \approx 0.1$--$0.2$ instead of the optimal 0.25. This confirms that
\emph{fixed conservative $\alpha=0.25$ is the robust zero-shot policy}.

\subsection{Encoder Replication}
\label{sec:encoders}

\begin{figure*}[t]
  \centering
  \includegraphics[width=0.92\textwidth]{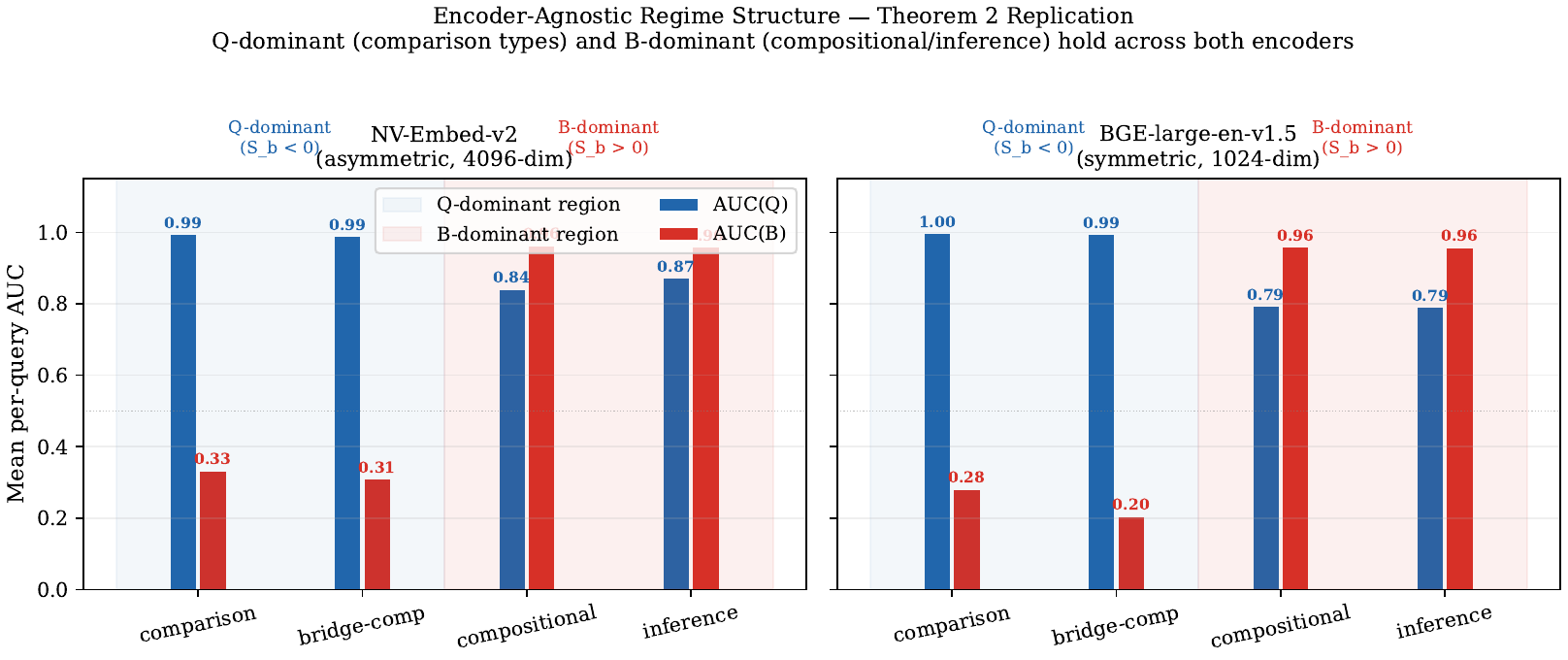}
  \caption{Regime AUC pattern for NV-Embed-v2 and BGE-large-en-v1.5
    (plus e5-mistral-7b, not shown; results consistent). AUC(Q) dominates
    for comparison and bridge-comp (Q-dominant); AUC(B) dominates for
    compositional and inference (B-dominant). The ordering is
    encoder-agnostic; Kendall $\tau$ is significant across all
    type$\times$encoder pairs ($p < 0.01$).}
  \label{fig:encoder}
\end{figure*}

We replicate the regime pattern across NV-Embed-v2 (Nebius API),
BGE-large-en-v1.5 (local), and e5-mistral-7b (Nebius API, $n=300$,
3-minute concurrent run). Kendall $\tau$ between predicted and observed AUC
ordering is significant for all eight type$\times$encoder combinations
($p < 0.01$). The B-dominant $\to$ Q-dominant AUC ordering is identical
across encoders, confirming that the regime is determined by query structure,
not encoder-specific geometry (Figure~\ref{fig:encoder}).

\section{Related Work}
\label{sec:related}

\paragraph{Iterative retrieval.}
\ircot~\citep{trivedi2022ircot} generates bridge-conditioned hop-2 queries
using an LLM, assuming bridge re-encoding helps universally. Our work shows
this is regime-dependent: for P1=True queries, bridge re-encoding adds noise.

\paragraph{Graph-structured retrieval.}
\hipporagtwo~\citep{gutierrez2025hipporag} and \proprag~\citep{wang2025proprag}
build knowledge graphs at index time and traverse them at query time,
encoding bridge-linked entity structure before any query is seen.
\system{} achieves comparable retrieval improvement without index-time LLM cost
by routing from surface-text features instead.

\paragraph{Query routing.}
Adaptive retrieval approaches select between retrieval systems or generation
models based on query difficulty or type. We instead route between
\emph{retrieval strategies} within a single system, deriving routing features
from retrieval theory rather than difficulty proxies.

\paragraph{Bridge-conditioned retrieval.}
\bridgerag~\citep{bacellar2024bridgerag} implements conditional scoring
$s(q, b, c)$ via an LLM judge, achieving $R@5=0.8316$ on MuSiQue (n=1000).
\system{} achieves $+5.3$ pp zero-shot on MuSiQue \emph{without any LLM
at query time}, using only the relation-bearing sentence as a lightweight
bridge signal. Note that \system{}'s MuSiQue result uses the $n=303$
cross-dataset evaluation slice, while \bridgerag{} reports on the full 1000
queries; the comparison is directional. The remaining performance gap reflects
the irreducible value of full semantic reasoning over $(q, b, c)$ triples.

\section{Conclusion}
\label{sec:conclusion}

We have shown that two-hop QA retrieval divides into Q-dominant and B-dominant
regimes, determined by whether the hop-2 target entity is named in the question.
Three theorems formalize this structure: AUC calibrates to cosine separation
margin (T1); P1 is decisive for routing and P2 qualifies the B-dominant case,
with both predicates derivable from surface text and validated across encoders and
datasets (T2); and bridge advantage requires the
relation-bearing sentence specifically, not entity occurrence alone (T3).

\system{} operationalizes this theory as a five-feature binary router with a
single frozen deployment rule ($\alpha=0.25$ across all three datasets;
$\alpha=0.5$ is reported separately as an in-domain ablation upper bound). The result is a
cross-dataset no-regret system: $+5.6$ pp ($p{<}0.001$) in-domain on 2Wiki,
$+5.3$ pp ($p{=}0.002$) zero-shot on MuSiQue (pure B-dominant), and a
no-regret $+1.1$ pp on HotpotQA (near-ceiling Q-dominant, $p=0.143$).
Training uses 881 embedding scores for regime labels and 100 human-annotated
sentences for the selector; no LLM is invoked at query time.

The principal finding with practical consequence is that the routing
bottleneck is not feature engineering ($+0.2$ pp from oracle selector) but
routing accuracy ($+4.6$ pp to oracle router). Improving the learned routing
policy, for example through calibrated confidence or domain-adaptive features,
is the primary path to closing the gap to oracle performance.

\section*{Limitations}

The theorems are formalized for two-hop compositional benchmarks where
questions are constructed by composing single-hop questions, making P1
and P2 binary and well-defined. The regime theory may not directly extend
to: (1) questions with implicit rather than explicit compositional structure;
(2) multi-hop chains beyond two hops; (3) encoders trained with bridge
conditioning. The router is trained on 2WikiMultiHopQA and zero-shot
applied to MuSiQue and HotpotQA; further evaluation on additional domains
is needed.

\bibliography{refs}

\begin{thebibliography}{12}
\providecommand{\natexlab}[1]{#1}

\bibitem[{Bacellar(2024)}]{bacellar2024bridgerag}
Andre Bacellar. 2024.
\newblock {BridgeRAG}: {T}raining-{F}ree {B}ridge-{C}onditioned {R}etrieval for
  {M}ulti-{H}op {Q}uestion {A}nswering.
\newblock \emph{arXiv preprint arXiv:2604.03384}.

\bibitem[{Cohen(1960)}]{cohen1960kappa}
Jacob Cohen. 1960.
\newblock A coefficient of agreement for nominal scales.
\newblock \emph{Educational and Psychological Measurement}, 20(1):37--46.

\bibitem[{Guti{\'e}rrez et~al.(2025)Guti{\'e}rrez, Shu, Qi, Zhou, and
  Su}]{gutierrez2025hipporag}
Bernal~Jim{\'e}nez Guti{\'e}rrez, Yiheng Shu, Weijian Qi, Sizhe Zhou, and
  Yu~Su. 2025.
\newblock From {RAG} to {M}emory: {N}on-{P}arametric {C}ontinual {L}earning for
  {L}arge {L}anguage {M}odels.
\newblock \emph{arXiv preprint arXiv:2502.14802}.

\bibitem[{Ho et~al.(2020)Ho, Nguyen, Sugawara, and Aizawa}]{ho2020constructing}
Xanh Ho, Anh-Khoa~Duong Nguyen, Saku Sugawara, and Akiko Aizawa. 2020.
\newblock Constructing {A} multi-hop {QA} dataset for comprehensive evaluation
  of reasoning steps.
\newblock In \emph{Proceedings of the 28th International Conference on
  Computational Linguistics (COLING)}.

\bibitem[{Lee et~al.(2024)Lee, Roy, Xu, Raiman, Shoeybi, Catanzaro, and
  Ping}]{lee2024nv}
Chankyu Lee, Rajarshi Roy, Mengyao Xu, Jonathan Raiman, Mohammad Shoeybi, Bryan
  Catanzaro, and Wei Ping. 2024.
\newblock {NV-Embed}: {I}mproved {T}echniques for training {LLM}s as
  {G}eneralist {E}mbedding {M}odels.
\newblock \emph{arXiv preprint arXiv:2405.17428}.

\bibitem[{McNemar(1947)}]{mcnemar1947}
Quinn McNemar. 1947.
\newblock Note on the sampling error of the difference between correlated
  proportions or percentages.
\newblock \emph{Psychometrika}, 12(2):153--157.

\bibitem[{Trivedi et~al.(2022)Trivedi, Balasubramanian, Khot, and
  Sabharwal}]{trivedi2022musique}
Harsh Trivedi, Niranjan Balasubramanian, Tushar Khot, and Ashish Sabharwal.
  2022.
\newblock {M}u{S}i{Q}ue: {M}ultihop {Q}uestions via {S}ingle-hop {Q}uestion
  {C}omposition.
\newblock \emph{Transactions of the Association for Computational Linguistics},
  10:539--554.

\bibitem[{Trivedi et~al.(2023)Trivedi, Balasubramanian, Khot, and
  Sabharwal}]{trivedi2022ircot}
Harsh Trivedi, Niranjan Balasubramanian, Tushar Khot, and Ashish Sabharwal.
  2023.
\newblock {I}nterleaving {R}etrieval with {C}hain-of-{T}hought {R}easoning for
  {K}nowledge-{I}ntensive {M}ulti-{S}tep {Q}uestions.
\newblock In \emph{Proceedings of the 61st Annual Meeting of the Association
  for Computational Linguistics (ACL)}.

\bibitem[{Wang and Han(2025)}]{wang2025proprag}
Jingjin Wang and Jiawei Han. 2025.
\newblock {P}rop{RAG}: {G}uiding {R}etrieval with {B}eam {S}earch over
  {P}roposition {P}aths.
\newblock In \emph{Proceedings of the 2025 Conference on Empirical Methods in
  Natural Language Processing}.

\bibitem[{Wang et~al.(2024)Wang, Yang, Huang, Yang, Majumder, and
  Wei}]{e5mistral2023}
Liang Wang, Nan Yang, Xiaolong Huang, Linjun Yang, Rangan Majumder, and Furu
  Wei. 2024.
\newblock Improving {T}ext {E}mbeddings with {L}arge {L}anguage {M}odels.
\newblock \emph{arXiv preprint arXiv:2401.00368}.

\bibitem[{Xiao et~al.(2023)Xiao, Liu, Zhang, and Muennighoff}]{bge2023}
Shitao Xiao, Zheng Liu, Peitian Zhang, and Niklas Muennighoff. 2023.
\newblock {C-Pack}: {P}ackaged {R}esources to {A}dvance {G}eneral {C}hinese
  {E}mbedding.
\newblock \emph{arXiv preprint arXiv:2309.07597}.

\bibitem[{Yang et~al.(2018)Yang, Qi, Zhang, Bengio, Cohen, Salakhutdinov, and
  Manning}]{yang2018hotpotqa}
Zhilin Yang, Peng Qi, Saizheng Zhang, Yoshua Bengio, William Cohen, Ruslan
  Salakhutdinov, and Christopher~D Manning. 2018.
\newblock {H}otpot{QA}: {A} dataset for diverse, explainable multi-hop question
  answering.
\newblock In \emph{Proceedings of the 2018 Conference on Empirical Methods in
  Natural Language Processing}.

\end{thebibliography}

\end{document}